\documentclass[aps,pra,reprint,amsmath,amssymb]
{revtex4-1}
\usepackage{graphicx}
\usepackage{bm}
\usepackage{hyperref}
\usepackage{color}
\usepackage{graphicx}
\usepackage{dcolumn}
\usepackage{braket}
\usepackage{soul} % added by Philippe
\usepackage{upgreek}
\usepackage{xcolor}

\graphicspath{{./Figures/}}

\begin{document}

\title{ Cooperative atomic emission from a line of atoms interacting with a resonant plane surface}

\author{M.\,O.~Ara\'ujo}
\affiliation{Departamento de F\'isica, Universidade Federal de Pernambuco, 50670-901, Recife, Pernambuco, Brazil}

\author{J.\,C.~de Aquino Carvalho}
\affiliation{Departamento de F\'isica, Universidade Federal de Pernambuco, 50670-901, Recife, Pernambuco, Brazil}

\author{Ph.\,W.~Courteille}
\affiliation{Instituto de F\'isica de S\~ao Carlos, Universidade de S\~ao Paulo, SP 13566-970 S\~ao Carlos, Brazil}

\author{A.~Laliotis}
\affiliation{Laboratoire de Physique des Lasers, UMR 7538 du CNRS, Universit\'e Sorbonne Paris Nord, F-93430 Villetaneuse, France}
 
\date{\today}

\begin{abstract} %Warning: around 200 words.
% Athanasios
Cooperative effects such as super- and subradiance can be observed in the fluorescence emitted by a system of $N$ atoms in vacuum, after interaction with a laser beam. In the vicinity of a dielectric or metallic surface, Casimir-Polder effects can modify collective atomic frequency shifts and decay rates. In this work, we study cooperative fluorescent emission next to resonant surfaces using the coupled dipoles model. We show that cooperative effects, expected in free space, are absent when the atoms are close to a surface whose polariton resonances coincide with the dominant atomic dipole coupling. In this case, cooperative effects are overshadowed by the very fast decay of the atomic fluorescence into surface modes. We illustrate our formalism and our results by considering a line of cesium $6D_{3/2}$ atoms in front of a sapphire surface. Finally, we propose the study of Cesium $6P_{3/2}$ atoms in front of a resonant metasurface as the most promising scenario for experimentally demonstrating the results of our study.
\end{abstract}

\maketitle

\section{Introduction}

%Cooperative effects
%Literature on Casimir-Polder effects
%Arruda's papers (cite them!)
%Cite recent papers
%Our work

When many atoms in vacuum interact with a coherent laser beam, the resonance fluorescence emitted by them may present cooperative effects, studied extensively following the seminal work of Dicke~\cite{Dicke1954}. Cooperative effects arise from quantum coherence created between the atoms if their relative distances are small, and the spontaneous emission has its decay rate enhanced. This enhancement, known as superradiance, was initially studied in the regime of many excited atoms in vapors~\cite{Gross1982}, but even if a single atom is excited, a situation known as single-photon superradiance~\cite{Scully2006, Courteille2010}, fast decay rates and frequency shifts %\st{can be observed} 
have been predicted.

In those cases, a system of $N$ two-level atoms with at most one quantum of excitation is prepared from laser excitation at low intensity and far detuning~\cite{Courteille2010}. The system evolves into a symmetric quantum state, called timed-Dicke state, and other $N-1$ anti-symmetric states, called subradiant states~\cite{Scully2007, Bienaime2011b}. The emission dynamics can be obtained from the so-called coupled-dipoles model~\cite{Svidzinsky2010, Bienaime2011}, which treats the atoms as oscillating dipoles interacting with the electromagnetic vacuum modes and a plane wave of incident light. %\Philippe{of incident light}. 
This model proved to be a versatile tool for modelling and predicting cooperative effects in several physical systems. Superradiance~\cite{Svidzinsky2008, Araujo2016, Roof2016, Kuraptsev2017}, subradiance~\cite{Bienaime2012, Guerin2016} and cooperative Lamb shift~\cite{Scully2009, Roof2016} were predicted and observed experimentally in the last decades, mainly in cold ensembles of atoms but also in hot vapors~\cite{Keaveney2012, Peyrot2018}, optical lattices~\cite{Olmos2010, Schilke2011} and Bose-Einstein condensates~\cite{Inouye1999}. Cooperative effects were predicted and/or observed also in other observables, such as the radiation pressure force~\cite{Bienaime2010} and intensity correlation functions~\cite{Das2008, Piovella2022}, and in non-linear phenomena~\cite{Oliveira2014}, spectral broadening~\cite{Jenkins2016} and resonance fluorescence~\cite{Pucci2017}. Also, the interplay between subradiance and radiation trapping~\cite{Chabe2014, Weiss2018} was investigated with cold atomic clouds, as well as the role played by finite temperature~\cite{Bromley2016, Weiss2019}. Applications are, among others, ultranarrow bandwidth %\Philippe{ultranarrow bandwidth} 
laser emission~\cite{Bohnet2012} and quantum information~\cite{Duan2001}.

%From a theoretical point of view, the emission dynamics is treated by two approaches: the coupled-dipole model and an eigenvalue approach. The former treats the atoms as oscillating dipoles and allows the calculation of the emitted fluorescence, among other observables~\cite{Bienaime2011}, both for low~\cite{Araujo2018} and high intensities~\cite{Cipris2021} of the driving laser. The latter is based on the calculation of the eigenvalues and eigenmodes of an effective Hamiltonian~\cite{Svidzinsky2010, Bellando2014, Guerin2017}. ????

%From a theoretical point of view, the emission dynamics can be treated by the coupled-dipole model~\cite{Svidzinsky2010, Bienaime2011}. The  atoms are treated as oscillating dipoles and allows the calculation of the emitted fluorescence, among other observables~\cite{Bienaime2011}, both for low~\cite{Araujo2018} and high intensities~\cite{Cipris2021} of the driving laser. An interaction Hamiltonian is written ????

Vacuum fluctuations are responsible for spontaneous emission and the displacement of energy levels of an isolated atom. These fluctuations also cause the Casimir force, in which two (massive) bodies, very close to each other, tend to mutually attract due to the decrease in energy density with respect to free space. An analogous effect is observed when one of the plates is replaced by a quantum object such as an atom. This interaction between atom and surface is known as Casimir-Polder interaction \cite{Casimir1948}. When the atom is located at distances smaller than the reduced wavelength of a transition ($\lambda/2\pi$), the Casimir-Polder interaction is in the near-field regime %(also known as van der Waals regime) 
~\cite{Casimir1948}. In this near-field regime, the atom-surface interaction can be viewed as an interaction between a fluctuating atomic dipole with its own surface-induced image. This dipole-dipole interaction has a potential of the type $C_3/h^3$, where $C_3$ is the van der Waals coefficient and $h$ is the distance between the surface and the atom. $C_3$ for a given state $\ket{i}$ is calculated as the sum of all allowed dipolar couplings $\ket{j}$ with frequency $\omega_{ij}$ (for emission $\omega_{ij}>0$ and for absorption $\omega_{ij}<0$). For an ideal conductor, we simply have $C_3 \propto \sum_j r(\omega_{ij}) |\bra{i}\mathbf{\mu}\ket{j}|^2$, where $\bra{i}\mathbf{\mu}\ket{j}$ is the dipole moment matrix element and $r(\omega_{ij})$ is the image coefficient. However, for an accurate interpretation of an interaction between atoms and a dielectric surface, it is necessary to insert the dielectric properties of the surface %permittivity properties of the dielectric 
through $r(\omega_{ij})$ at zero temperature given by \cite{Fichet1995,Gorza2006},
\begin{equation}
\label{reflexao}
    r(\omega_{ij}) = \frac{2}{\pi}\int_{0}^{\infty}S(i\xi)\frac{\omega_{ij}}{\xi^2 + \omega_{ij}^2}d\xi - 2\mathrm{Re}\left[S(\omega_{ij})\right]
\end{equation}
where the first term can be seen as a renormalization of the vacuum due to the presence of the surface and the second term resembles the interaction of a classical dipole with its own field reflected at the surface \cite{Wylie1984}. For a perfect conductor, $r(\omega_{ij})=1$. In Eq.~\eqref{reflexao}, $S(\omega)=\frac{\varepsilon(\omega)-1}{\varepsilon(\omega)+1}$ is the surface response, with $\varepsilon(\omega)$ the complex permittivity of the surface. Note that there %might 
may exist a frequency $\omega$ where $\varepsilon(\omega)=-1$ and, consequently, where $S(\omega)$ diverges. This $\omega$ is known as surface polariton frequency. When an atomic transition frequency $\omega_{ij}$ coincides with a surface polariton frequency, we say that we are in the atom-surface resonant interaction regime. These resonant effects can make the atom-surface interaction go from attractive to repulsive as well as influence the lifetime of excited states %(see for example the references \cite{Failache1999,Failache2002}).
(see, e.g., refs. \cite{Failache1999,Failache2002} for interactions between a Cs vapor and sapphire). Temperature effects may also have an impact on the atom-surface interaction~\cite{Gorza2006, Barton1997, Laliotis2015} and have been observed in both, non-resonant \cite{Laliotis2014} and resonant cases \cite{Joao2023}.
%The cooperative environment of an atom can also be modified by adding a metallic or dielectric surface nearby. This effect, refereed as Casimir-Polder effect, has to do with the modification of the vacuum electromagnetic modes by the surface, and the interaction between atom and this modified vacuum leads to frequency shifts in the atomic states proportional to $1/h^3$, where $h$ is the distance between atom and surface. Theoretical and experimental work were performed in Cs hot vapors interacting with sapphire, YAG and glass surfaces~\cite{Ducloy1991, Gorza2006, Laliotis2014}. Casimir-Polder effects were also observed in atomic beams deflection~\cite{Sandoghdar1992} and also in a cloud of cold atoms interacting with a dielectric prism~\cite{Bender2010}. If a surface presents a resonance wavelength $\lambda_s$ comparable to an atomic resonance frequency $\lambda$, the relative dieletric permittivity of the surface is $\epsilon=-1$, a condition called surface polariton, and the atom looses its excitation to a surface mode~\cite{Failache1999,Failache2002}. \Philippe{I find this not that obvious.} \textcolor{blue}{To improve this whole paragraph (Joao or Athanasios)}

Published theoretical work focuses on the calculation of corrections to the energies and decay rates from second-order perturbation theory in the interaction Hamiltonian between atom and surface for a single atom~\cite{Wylie1984}. Another approach consists in determining the Green function in terms of the Fresnel coefficients from the interaction between atom and the surface-modified electric field~\cite{Tomas1995} for one and two atoms embedded in a multilayer dielectric~\cite{Tomas1995, Dung2002}. Interplay between cooperative effects and Casimir-Polder interactions was also predicted theoretically. For two Rb atoms, cooperative decay rates and shifts were predicted to be modified due to Fano resonances, when the atoms are near a nanosphere~\cite{Arruda2020}. For $N>2$ atoms near a surface, theoretical work predicted changes in superradiance~\cite{Fuchs2018} and of the Casimir-Polder force in the regime of many excitations~\cite{Fuchs2018, Sinha2018}. An approach based on the calculation of frequency shifts and decay rates from the Green matrix was done for a line of two to twenty atoms interacting with an Ag surface~\cite{Jones2018}. However, work on atoms interacting with surfaces are still scarce, specially when atoms and surface are in resonance.

The control of cooperative effects, in particular, subradiance, is interesting for applications in quantum information, such as fast readout~\cite{Scully2015} and entangled state generation~\cite{Santos2022}. In this work, we combine the coupled-dipole model, which predicts cooperative effects as super- and subradiance, with the presence of plane surfaces, and we study the possibility of coupling super- and subradiant modes with surface-induced resonances. Contrary to refs.~\cite{Sinha2018, Fuchs2018}, which predicted an enhancement of the Casimir-Polder forces for $N \gg 1$ excited atoms close to a surface, here we study the linear-optics regime, i.e., the single-excitation limit. Our formalism is illustrated by considering a sapphire surface interacting with a line of $N$ Cs ($6D_{3/2}$) atoms, presenting a mid-infrared coupling ($6D_{3/2} \rightarrow 7P_{1/2}$ at $12.15$ $\mu$m) resonant with the sapphire polariton at $12.15~\upmu$m, studied experimentally since the $90$'s. By evaluating the emitted fluorescence from the coupled dipoles model after the system has reached a steady state, we show that super- and subradiance vanish %are spoiled
due to the surface polaritons, and the fluorescence of the whole system decays very fast as if individual (totally uncorrelated) atoms were placed close to the surface. In the absence of surface polaritons, i.e., for interaction of Cs with glass or a metallic surface, cooperative effects remain present but are modified %are not \textcolor{blue}{spoiled, but simply modified, 
by the presence of the surface. For an experimental demonstration of the above effects, we propose studies of the $6P_{3/2}\rightarrow 6S_{1/2}$ decay channel at $852$ nm (better approximated by a two-level system), coupled with a resonant metasurface.

This paper is organized as follows. In section \ref{sec_model}, we review the approach for cooperative effects and surface interactions for $N$ atoms (subsections \ref{subsec_model_a} and \ref{subsec_model_b}) and then present a modified coupled-dipoles model taking into account surface effects (\ref{subsec_model_c}), in order to evaluate the decay dynamics of the atoms. In section \ref{sec_methods} we present our simulation methods, and the main results are presented in section \ref{sec_results}. We point out the general importance of our formalism in section~\ref{sec_discussion}. We make our concluding remarks in section \ref{sec_conclusion}.

%\vspace{2cm}
%\textcolor{blue}{Remark: Something has crashed when I was editing and compiling section II (theoretical model). Fortunately, I saved all the text and I will put it back later.}
%\vspace{2cm}

%------------------------------------------------------------
\section{Theoretical model}

\label{sec_model}
% CD model: calculations from Olmos and Cipris

\subsection{Atoms in free space and close to a plane surface}
\label{subsec_model_a}

%Even though it is possible to derive the equations describing the evolution of coupled dipoles in the low excitation limit from classical equations, a quantum formalism is often conveniently used.

\begin{figure}[h]
    \centering
    \includegraphics[scale=1]{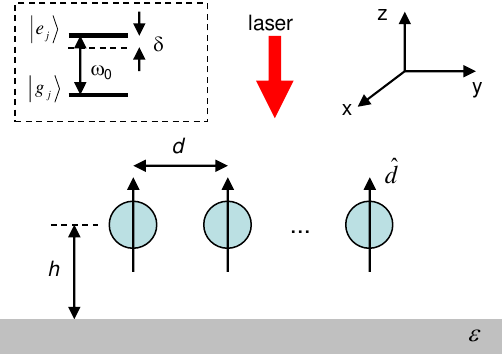}
    \caption{Scheme of the physical system. A line of $N$ identical two-level atoms is placed at identical distances $h$ from a planar surface lying on the semi-infinite $z\leq0$ plane. The atoms are distributed with equal distances $d$ along the $y$-axis and have their dipoles oriented in the direction $\mathbf{\hat{d}}=\mathbf{\hat{z}}$, i.e., perpendicularly to the surface. A laser beam of frequency $\omega$ pointing into the $-\mathbf{\hat{z}}$ direction drives the atoms. Inset: Scheme of the two atomic levels. %\Philippe{What is $k$ or $k_0$? When specifying distances, better choose quantities with units of length. Scaled distances only are nice in formulas. Eventually, remove $k_0$ from the figure!}
    }
    \label{fig1}
\end{figure}

The formalism used here is the same as presented in \cite{Tomas1995, Dung2002, Jones2018}, which considers dipoles in vacuum close to a planar surface, in such a way that the vacuum and the surface form a two-layer medium. We consider a line of $N$ identical two-level atoms with resonance frequency $\omega_0$, transition wavelength $\lambda$ and same dipole orientations $\mathbf{\hat{d}}$ (Fig.~\ref{fig1}) \cite{footnote}. The atomic levels of the atom $j$ are denoted by $\left| g_j \right\rangle $ ($\left| e_j \right\rangle$) for the ground (excited) state, with $j=1,...,N$. %\st{Although light is described as a vector field, it is not necessary to include three excited states in the model.} 
All atoms are fixed at positions $\textbf{r}_a=(x_a,y_a,z_a)$ and placed at a distance $h$ from the surface of a plane dielectric occupying the half space $z\leq 0$. The other half space $z\geq 0$ is vacuum. A monochromatic plane wave of frequency $\omega$ and detuning $\Delta=\omega-\omega_0$ is incident along the $z$ axis and illuminates all atoms.

% Fig1: setup

%\textcolor{blue}{Remark: I checked all analytical calculations.}

The Heisenberg-Liouville equation for the Pauli spin-down operator acting on atom \textit{a} reads
\begin{equation}
	\dot{\hat\sigma}_a^- = -\tfrac{i}{\hbar}[\hat\sigma_a^-,\hat H]+\mathfrak{L}^\dagger[\hat\sigma_a^-]
    \label{HeiLiou}
\end{equation}
where $\hat\sigma_a^-=\left|g_a\right>\left<e_a\right|$ is the atomic operator. The Hamiltonian and dissipation terms are
%\begin{align}
%	\hat H & = \tfrac{\hbar}{2}\sum_b[\Omega(\mathbf{r}_b)\hat\sigma_b^++h.c.]-\hbar\sum_{a,b}V_{ba}\hat\sigma_b^+\hat\sigma_a^-\\
%	\mathfrak{L}^\dagger[\hat A] & = \tfrac{1}{2}\sum_{a,b}\Gamma_{ab}\left(2\hat\sigma_a^+\hat A\hat\sigma_b^--\hat\sigma_a^
%		+\hat\sigma_b^-\hat A-\hat A\hat\sigma_a^+\hat\sigma_b^-\right)\nonumber
%\end{align}
\begin{eqnarray}
\hat H &=& -\frac{\hbar\Delta}{2}\sum_b \hat\sigma_b^z + \tfrac{\hbar}{2}\sum_b[\Omega(\mathbf{r}_b)\hat\sigma_b^++h.c.]-\nonumber\\
&-& \hbar\sum_{a,b}V_{ab}\hat\sigma_a^+\hat\sigma_b^-\\
\mathfrak{L}^\dagger[\hat A] &=& \tfrac{1}{2}\sum_{a,b}\Gamma_{ab}\left(2\hat\sigma_a^+\hat A\hat\sigma_b^--\hat\sigma_a^
		+\hat\sigma_b^-\hat A-\hat A\hat\sigma_a^+\hat\sigma_b^-\right)~~~~
\end{eqnarray}
where $\hat\sigma_a^z\equiv \hat\sigma_a^+\hat\sigma_a^- - \hat\sigma_a^-\hat\sigma_a^+$ and $\Omega(\mathbf{r})$ denotes the Rabi frequency generated by a classical incident laser beam at position $\mathbf{r}$. Collective shifts and decay rates are given by 

%\textcolor{black!50!white}{
%\begin{align}
%    V_{a\neq b} & = -\tfrac{3\lambda\Gamma}{2}\mathbf{\hat d}^*\text{Re~}G^0(\mathbf{r}_a,\mathbf{r}_b,\omega_0)\mathbf{\hat d} ~~~~~\text{and}~~~~~ V_{aa} = \delta\nonumber\\
%	\Gamma_{a\neq b} & = -3\lambda\Gamma\mathbf{\hat d}^*\text{Im~}G^0(\mathbf{r}_a,\mathbf{r}_b,\omega_0)\mathbf{\hat d} ~~~~~\text{and}~~~~~ \Gamma_{aa} = \Gamma\nonumber
%\end{align}
%}

% Modifiquei 3Gamma/2 --> Gamma
\begin{subequations}
\label{Vab_Gab}
\begin{eqnarray}
V_{ab} &=& \lambda\Gamma\mathbf{\hat{d}}^\ast\text{Re~}G(\mathbf{r}_a,\mathbf{r}_b,\omega_0)\mathbf{\hat{d}} \\
\Gamma_{ab} &=& 2\lambda\Gamma \mathbf{\hat{d}}^\ast\text{Im~}G(\mathbf{r}_a,\mathbf{r}_b,\omega_0)\mathbf{\hat{d}}
%\Gamma_{ab} &=& -3\lambda\Gamma\mathbf{\hat{d}}^\ast\text{Im~}G^0(\mathbf{r}_a,\mathbf{r}_b,\omega_0) \mathbf{\hat d}
\end{eqnarray}
\end{subequations}
%\begin{subequations}
%\label{Vab_Gab}
%\begin{eqnarray}
%V_{ab} &=& \frac{3\lambda\Gamma}{2}\mathbf{\hat{d}}^\ast\text{Re~}G(\mathbf{r}_a,\mathbf{r}_b,\omega_0)\mathbf{\hat{d}} \\
%\Gamma_{ab} &=& 3\lambda\Gamma \mathbf{\hat{d}}^\ast\text{Im~}G(\mathbf{r}_a,\mathbf{r}_b,\omega_0)\mathbf{\hat{d}}
%\Gamma_{ab} &=& -3\lambda\Gamma\mathbf{\hat{d}}^\ast\text{Im~}G^0(\mathbf{r}_a,\mathbf{r}_b,\omega_0) \mathbf{\hat d}
%\end{eqnarray}
%\end{subequations}

In Eqs.~\eqref{Vab_Gab}, $G$ is the Green tensor describing the interaction between two atoms \textit{a} and \textit{b}, and $\Gamma=\omega_0^3 d_\text{eg}^2/(3\pi c^3\varepsilon_0\hbar)$ is the free-space single atom decay rate, $d_\text{eg}$ is the matrix dipole element between the states $|g\rangle$ and $|e\rangle$. %\Philippe{and $d_\text{eg}$ is the matrix dipole element between the states $|g\rangle$ and $|e\rangle$}. 
As in~\cite{Jones2018}, we assume all dipoles to have the same orientation $\mathbf{\hat{d}}$. With this, Eq.~\eqref{HeiLiou} becomes 
%\textcolor{blue}{PS:Who is $\delta$ in equation below? If Eqs. \ref{HeiLiou} are derived for free space, we should have $\delta=0$...}
%\begin{align}
%    \label{HeiLiou2}
%	\dot{\hat\sigma}_a^- & = \tfrac{i}{2}\Omega(\mathbf{r}_a)\hat\sigma_a^z-i\delta\hat\sigma_a^z\hat\sigma_a^-+\tfrac{\Gamma}{2}\hat\sigma_a^z\hat\sigma_a^-\\
%    & -i\sum_{b_{\neq a}}^N
%	\Delta_{ab}\hat\sigma_a^z\hat\sigma_b^-+\sum_{b_{\neq a}}^N\tfrac{\Gamma_{ab}}{2}\hat\sigma_a^z\hat\sigma_b^-\nonumber 
%\end{align}
\begin{align}
    \label{HeiLiou2}
	\dot{\hat\sigma}_a^- & = i\Delta\hat\sigma_a^- + \tfrac{i}{2}\Omega(\mathbf{r}_a)\hat\sigma_a^z-i\delta\hat\sigma_a^z\hat\sigma_a^-+\tfrac{\Gamma_{tot}}{2}\hat\sigma_a^z\hat\sigma_a^-\\
    & -i\sum_{b_\neq a}^N
	V_{ab}\hat\sigma_a^z\hat\sigma_b^-+\sum_{b_\neq a}^N\tfrac{\Gamma_{ab}}{2}\hat\sigma_a^z\hat\sigma_b^-\nonumber 
\end{align}
where $\delta\equiv V_{aa}$ and $\Gamma_{tot}\equiv\Gamma_{aa}$. This equation will be the starting point used in Sec.~\ref{subsec_model_c} for the derivation of the coupled dipoles model, which is the basis for our numerical simulations.

Eq.~\eqref{Vab_Gab} has been derived for free space. The presence of a surface can now be accounted for by simply adding to the free space Green function a scattering Green function%}
\begin{equation}
    G = G^0+G^\text{R}
\end{equation}
where $G^0$ is the Green tensor in free space and $G^\text{R}$ is the scattering term taking into account the surface. This decomposition allows us to write
\begin{subequations}
\label{Vab_Gab_total}
\begin{eqnarray}
    V_{ab}&=& V_{ab}^0+V_{ab}^\text{R} \label{Vab_decomposed} \\
    \Gamma_{ab}&=& \Gamma_{ab}^0 + \Gamma_{ab}^\text{R} 
    \label{Gab_decomposed}
\end{eqnarray}
\end{subequations}
because of Eqs.~\eqref{Vab_Gab}.

\subsection{Green tensor for free space and surface}
\label{subsec_model_b}

The Green tensor for free space, $G^0$, is given in~\cite{Jones2018,Buhmann2012}. Eqs.~\eqref{Vab_Gab} with $G=G^0$ then lead to~\cite{Jones2018}

% Modifiquei 3Gamma/2 --> Gamma
\begin{subequations}
\label{Vab0_Gab0}
\begin{eqnarray}
V_{ab}^0&=& \dfrac{\Gamma}{2} \left[ (1-(\mathbf{\hat{d}} \cdot\mathbf{\hat{r}}_{ab})^2) \dfrac{\cos\kappa_{ab}}{\kappa_{ab}} - \right. \nonumber \\
&-& \left. (1-3(\mathbf{\hat{d}}\cdot\mathbf{\hat{r}}_{ab})^2) \left( \dfrac{\sin\kappa_{ab}}{\kappa_{ab}^2} + \dfrac{\cos \kappa_{ab}}{\kappa_{ab}^3} \right) \right] \label{shift_vectorial} \\
\Gamma_{ab}^0&=& \Gamma \left[ (1-(\mathbf{\hat{d}}\cdot\mathbf{\hat{r}}_{ab})^2) \dfrac{\sin\kappa_{ab}}{\kappa_{ab}} + \right. \nonumber \\
&+& \left. (1-3(\mathbf{\hat{d}}\cdot\mathbf{\hat{r}}_{ab})^2) \left( \dfrac{\cos\kappa_{ab}}{\kappa_{ab}^2} - \dfrac{\sin \kappa_{ab}}{\kappa_{ab}^3} \right) \right] \label{decay_vectorial}
\end{eqnarray}
\end{subequations}
%\begin{subequations}
%\label{Vab0_Gab0}
%\begin{eqnarray}
%V_{ab}^0&=& \dfrac{3\Gamma}{4} \left[ (1-(\mathbf{\hat{d}} \cdot\mathbf{\hat{r}}_{ab})^2) \dfrac{\cos\kappa_{ab}}{\kappa_{ab}} - \right. \nonumber \\
%&-& \left. (1-3(\mathbf{\hat{d}}\cdot\mathbf{\hat{r}}_{ab})^2) \left( \dfrac{\sin\kappa_{ab}}{\kappa_{ab}^2} + \dfrac{\cos \kappa_{ab}}{\kappa_{ab}^3} \right) \right] \label{shift_vectorial} \\
%\Gamma_{ab}^0&=& \dfrac{3\Gamma}{2} \left[ (1-(\mathbf{\hat{d}}\cdot\mathbf{\hat{r}}_{ab})^2) \dfrac{\sin\kappa_{ab}}{\kappa_{ab}} + \right. \nonumber \\
%&+& \left. (1-3(\mathbf{\hat{d}}\cdot\mathbf{\hat{r}}_{ab})^2) \left( \dfrac{\cos\kappa_{ab}}{\kappa_{ab}^2} - \dfrac{\sin \kappa_{ab}}{\kappa_{ab}^3} \right) \right] \label{decay_vectorial}
%\end{eqnarray}
%\end{subequations}
\\*
where $\kappa_{ab}=kr_{ab}$,  $r_{ab}=|\textbf{r}_{ab}|=|\textbf{r}_a - \textbf{r}_b|$ is the relative distance between the atoms $a$ and $b$ and $\mathbf{\hat{r}}_{ab}$ is the unit vector along the direction of $\textbf{r}_{ab}$. As pointed out by \cite{Jones2018}, the expressions~\eqref{Vab0_Gab0} yield the frequency shifts and decay rates of the excited energy levels %in the so-called vectorial \Philippe{approximation (I thought vectorial is exact?)} 
obtained in previous work~\cite{Lehmberg1970, Bienaime2013, Cipris2021}. They lead to cooperative effects as, e.g., superradiance~\cite{Araujo2016, Roof2016}, subradiance~\cite{Guerin2016} and cooperative Lamb shift~\cite{Roof2016}. For $a=b$, we have $V_{aa}^0=0$ and $\Gamma_{aa}^0=\Gamma$.

The Green tensor $G^\text{R}$ for a surface is given in refs.~\cite{Dung2002, Jones2018} for atoms in a multilayer dielectric, and takes into account the reflection of vacuum modes and evanescent modes created by %the surface. %\st{In the case}
each layer. The present case of a vacuum-dielectric interface can be seen as a two-layer dielectric, so the equations of ref.~\cite{Jones2018} simplify and read
%\footnote{The equations presented in this section for $G^R$ terms are the same given in the Appendix B of \cite{Jones2018} with the simplifications $h=0$, $r_{+}^q=0$ and $D^q=1$ in their notation.}
\begin{equation}
    G^\text{R}(\textbf{r}_a,\textbf{r}_b,\omega_0)=\dfrac{i}{4\pi} \int_0^{\infty} dk_{\rho} \dfrac{k_{\rho}}{k_z}\left(G^\text{s}-\dfrac{k_z^2}{k^2} G^\text{p}\right)
    \label{GR}
\end{equation}
where $G^\text{s}$ and $G^\text{p}$ are given by
\begin{subequations}
    \label{Gs_Gp}
    \begin{eqnarray}
        \label{Gs}
        G^\text{s}&=&\dfrac{r^\text{s}e^{2ik_zh}}{2} 
        \begin{pmatrix}
        J_0-J_2 & 0 & 0 \\ 0 & J_0+J_2 & 0 \\ 0 & 0 & 0
        \end{pmatrix}\\
        G^\text{p}&=&\dfrac{r^\text{p}e^{2ik_zh}}{2} 
        \begin{pmatrix}
        J_0+J_2 & 0 & 0 \\ 0 & J_0+J_2 & \dfrac{2ik_\rho}{k_z}J_1\sin\phi \\ 0 & -\dfrac{2ik_\rho}{k_z}J_1\sin\phi & -\dfrac{2k_\rho^2}{k_z^2}J_0 
        \end{pmatrix} \nonumber\\
        \label{Gp}
        &&
    \end{eqnarray}
\end{subequations}

In Eqs.~\eqref{Gs} and \eqref{Gp}, $k=2\pi/\lambda=\omega_0/c$, $k_z=\sqrt{k^2-k_\rho^2}$ [with $\mathrm{Re}(k_z)>0$, $\mathrm{Im}(k_z)>0$ \cite{Tomas1995}], $J_n\equiv J_n(k_\rho|y_{ab}|)$ is the Bessel function of order $n$, and $\sin\phi=y_{ab}/|y_{ab}|$, which gives $\sin\phi=+1$ ($\sin\phi=-1$) for $a<b$ ($a>b$). The quantities $r^q$ with $q=\{\text{s},\text{p}\}$, are the Fresnel coefficients and are given by
\begin{subequations}
    \label{fresnel_coeffs}
    \begin{eqnarray}
        r^s &=& \dfrac{k_z-k_{IIz}}{k_z+k_{IIz}} \\
        r^p &=& \dfrac{\epsilon(\omega)k_z-k_{IIz}}{\epsilon(\omega)k_z+k_{IIz}}
    \end{eqnarray}
\end{subequations}
with $k_{IIz}=\sqrt{\epsilon(\omega)k^2-k_\rho^2}$, [$\mathrm{Re}(k_{IIz})>0$ and $\mathrm{Im}(k_{IIz})>0$].

The surface shifts $V_{ab}^\text{R}$ and decay rates $\Gamma_{ab}^\text{R}$ are given by Eqs.~\eqref{Vab_Gab} with $G=G^\text{R}$. For $a=b$, we have the Casimir-Polder shift $\delta\equiv V_{aa}^\text{R}$ and the surface-induced decay rate $\Gamma_z\equiv \Gamma_{aa}^\text{R}$ in a such way that the total shift and decay rate are $V_{aa}=\delta$ and $\Gamma_{aa}=\Gamma+\Gamma_z$ (Eqs. \ref{Vab_Gab_total} with $a=b$). % with $V_{aa}^\text{R}\equiv\delta_\text{CP}$ and $\Gamma_{aa}^\text{R}\equiv\Gamma_z$. \Philippe{$\delta_\text{CP}$ and $\Gamma_z$ have not been introduced. Invert the definitions?} 
As pointed out by~\cite{Jones2018}, the total $V_{ab}$ and $\Gamma_{ab}$ contain the cooperative effects in the sense that diagonal terms ($a=b$) contain the shifts and decay rates for a single atom modified by the surface, whereas the off-diagonal terms ($a\neq b$) measure the strength of the cooperative effects due to the coupling between atoms and surface.

For a single atom ($N=1$) and a surface, we have $V_{a\neq b}^\text{R}=\Gamma_{a\neq b}^\text{R}=0$, and then we have $V_{aa}=\delta$ and $\Gamma_{aa}=\Gamma+\Gamma_z$, due to the interaction with the surface. Also, it can be shown~\cite{Dung2002} that for $kh\ll 1$, the matrix terms of $G^\text{R}$ become proportional to
\begin{equation}
\label{GR_approx}
    G_{ij}^\text{R}(\omega_0)\propto \frac{S(\omega_0)}{h^3}
\end{equation}
for $i,j=\{x,y,z\}$, where $S(\omega)=\tfrac{\varepsilon(\omega)-1}{\varepsilon(\omega)+1}$ and $\varepsilon(\omega)$ is the complex electric permittivity of the surface. Eq.~\eqref{GR_approx} is in agreement with Eq.~\eqref{reflexao} obtained in~\cite{Wylie1984}. Note that if $\varepsilon\approx -1$ for a certain atomic transition frequency $\omega$, a condition known as surface polariton, $S$ diverges, as well as  %and consequently 
$G_{ij}^\text{R}(\omega_0)$. This resonant enhancement effect was used in Cs hot vapors interacting with a sapphire surface to turn an atom-surface interaction from attractive to repulsive \cite{Failache1999} or cause %\Philippe{cause} 
the atomic emission to be absorbed by the surface \cite{Failache2002}. As we will see in the next subsection, the fluorescence emitted by the system depends on all these terms, and the effect of this enhancement for weak excitations is to modify the decay dynamics and to extinguish cooperative effects.

\subsection{Decay dynamics with surface}
\label{subsec_model_c}

Now, we assume low atomic excitation, $\langle\hat\sigma_a^z\rangle\simeq-1$ (low-energy Dicke state), that is, most atoms are in the ground state. Then, we may neglect correlations ($\langle\hat\sigma_a^z\hat\sigma_a^-\rangle=\langle\hat\sigma_a^z\rangle\langle\hat\sigma_a^-\rangle$) and find from Eq.~\eqref{HeiLiou2}, with $\beta_a\equiv\langle\hat\sigma_a^-\rangle$,
\begin{equation}
	\dot{\beta}_a \simeq \left(\imath\Delta_\text{tot}-\tfrac{\Gamma_\text{tot}}{2}\right)\beta_a-\tfrac{\imath}{2}\Omega(\mathbf{r}_a)+\sum_{b_{\neq a}}^N
	\left(\imath V_{ab}-\tfrac{\Gamma_{ab}}{2}\right)\beta_b
\label{betas_eq}
\end{equation}
\\*
where $\Delta_\text{tot}=\Delta+\delta$ and $\Gamma_\text{tot}=\Gamma+\Gamma_z$. In the absence of a surface, i.e., $G=G^0$, the dynamics of the $N$ atoms interacting with vacuum and a light field was discussed extensively~\cite{Bienaime2013}. Driving is weak for large detunings $\Delta$ and small Rabi frequencies $\Omega$, and under these conditions the system admits two states: the ground state $|G\rangle=|g_1 ... g_N\rangle$ (i.e., all $N$ atoms are in the state $|g_j\rangle$) and $N$ single-excitation states $|j\rangle=|g_1...e_j...g_N\rangle$ (i.e., the atom $j$ is in the state $|e_j\rangle$ and the other $N-1$ atoms are in the state $|g_j\rangle$).

Assuming that driven by a laser field the system reached a steady-state and then the laser is turned off, %\Philippe{driven by a laser field the system reached a steady-state and then the laser is turned off}
%the system is in the initial state $|\Psi(0)\rangle=|G\rangle$ and, having reached a steady state, the incoming laser field is turned off,
i.e., $\Omega=0$, the total fluorescence emitted by the atoms can be evaluated from~\cite{Araujo2018}
\begin{equation}
    P(t) \propto -\dfrac{d}{dt}\sum_{j=1}^N|\beta_j(t)|^2
    \label{Pt}
\end{equation}

Signatures of cooperative effects in the fluorescence were predicted theoretically~\cite{Bienaime2013} and observed experimentally~\cite{Guerin2016, Araujo2016, Roof2016} in the last decade for atoms distributed in free space.

For a single atom in free space Eqs.~\eqref{betas_eq} and \eqref{Pt} reduce to the standard fluorescence with natural exponential decay rate $\Gamma$. For a single atom near a surface Eqs.~\eqref{betas_eq} and \eqref{Pt} reproduce the decay dynamics with surface effects. Finally, for $N$ atoms near a surface, both surface and cooperative effects are present, and we show in the next section, that the cooperative effects are suppressed due to the surface modes when the surface is in resonance with the atomic transition, i.e., when $\varepsilon=-1$. For $\varepsilon\neq -1$, cooperative effects are slightly modified. 

%\st{
%%\subsection{
%D. Comparison with the Casimir-Polder effect for a single atom and surface%}
%}

%\st{Maybe it's worth to discuss a comparison between equations from the Casimir-Polder literature and the equations presented here. TO SEE THIS LATER.}

%-----------------------------------------------------------
\section{Simulation methods}\label{sec_methods}

\subsection{Casimir-Polder shift for Cs atom and sapphire}

Our simulation method is similar to the one described in~\cite{Araujo2018}. As in \cite{Jones2018}, we simulate the simplest case of a line of $N$ atoms distributed along the $y$-axis, equally spaced by a distance $d$, located at the same distance $h$ from a planar surface (see Fig.~\ref{fig1}), and having same dipole orientations $\mathbf{\hat{d}}$. That is, %\st{For given $N$, $kd$ and $kh$, we have} \Philippe{That is,} 
$x_a=0$ and $z_a=h$ for all atoms. The incoming laser comes from the direction $-\mathbf{\hat{z}}$, i.e., it illuminates all atoms equally. Then, for a given dipole orientation $\mathbf{\hat d}$, we evaluate $V_{ab}^0$ and $\Gamma_{ab}^0$ from Eqs.~\eqref{Vab0_Gab0}. We also compute the matrix terms of $G^\text{R}$ from Eqs.~(\ref{GR}-\ref{Gs_Gp}) and use them to evaluate $V_{ab}^\text{R}$ and $\Gamma_{ab}^\text{R}$ from Eqs.~\eqref{Vab_Gab}. Finally, we solve numerically the Eqs.~\eqref{betas_eq} governing the evolutions of the $\beta_a(t)$ for a given $\Delta$, and compute the total fluorescence $P(t)$ from Eq.~\eqref{Pt}. The fluorescence is normalized by its maximum value $P(0)$, which occurs at $t=0$ when steady state is reached and the incident driving field is turned off.

For atoms in free space, there is no shift in the excited state, so the frequency detuning of the laser $\Delta$ is taken with respect to the excited state energy. However, for atoms in the presence of a surface, as the excited level will be shifted by $\delta$, the laser frequency is set in a such way as to maintain the detuning from the shifted level, $\Delta_\text{tot}=\Delta+\delta$. %\Philippe{For atoms in free space, there is no shift in the excited state, so the frequency detuning of the laser $\Delta$ is taken with respect to the excited state energy. However, for atoms in the presence of a surface, as the excited level will be shifted by $\delta$, the laser frequency is set in a such way as maintain the detuning from the shifted level, $\Delta_\text{tot}=\Delta+\delta$.} 
%\st{For atoms in free space, there is no shift in the excited state, so the frequency detuning $\Delta$ is taken with respect to the excited state energy. However, for atoms in the presence of a surface, as the excited level will be shifted by $\delta$, the detuning $\Delta$ is set in a such way that the total detuning $\Delta_\text{tot}'=\Delta-\delta$ is equal to the detuning $\Delta$ in free space. As an example, if $\delta=+65\Gamma$ for a given surface and $\Delta=10\Gamma$, we set $\Delta=10\Gamma$ for atoms in free space and $\Delta=75\Gamma$ for atoms near the surface, because the net detuning in the latter will be $\Delta_\text{tot}'=75\Gamma-65\Gamma=10\Gamma=\Delta$.}

\begin{figure}[h]
    \centering
    \includegraphics[scale=0.8]{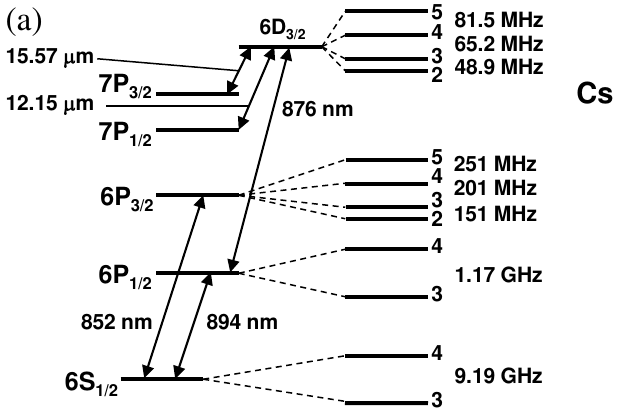} % New
    \includegraphics[scale=1]{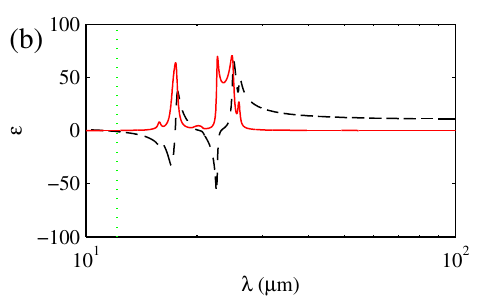}
    \includegraphics[scale=1]{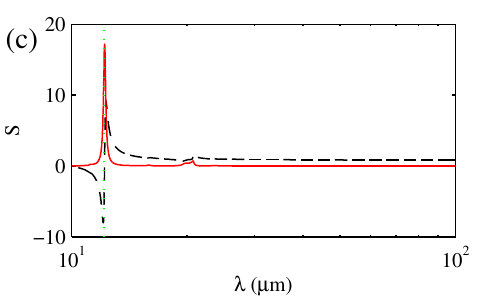}
    \caption{(a) Schematic of Cs levels. Real (dashed black) and imaginary (full red) parts of (b) the relative dielectric constant $\varepsilon$ and (c) $S=(\varepsilon-1)/(\varepsilon+1)$ for sapphire. The dotted green lines indicate the value $\lambda=12.15\,\upmu$m, for which $\varepsilon=-0.95 \approx -1$ and $S=\infty$.}
    \label{fig2}
\end{figure}

%We are interested in simulating Cs atoms close to a sapphire surface. 
As already mentioned, we apply our formalism to Cs atoms and a sapphire surface. Cesium has the transition decay $6D_{3/2}\rightarrow 7P_{1/2}$ with wavelength $\lambda=12.15\,\upmu$m (see Fig.~\ref{fig2}a), while sapphire has a resonance at approximately $\lambda_\text{s} \sim 12\,\upmu$m \cite{Fichet1995}, for which $\mathrm{Re~}\varepsilon\approx -1$ (Figs.~\ref{fig2}b-c), that is, we have a resonant atom-surface interaction. The choice of this transition and this surface is based on the fact that Casimir-Polder interaction between Cs and sapphire was extensively studied~\cite{Laliotis2021}, and sapphire has a full theoretical model for its dielectric constant~\cite{Laliotis2015}, so we consider it is important to have at hand a working theoretical tool. The Cesium transition has a natural decay rate of $\Gamma=2\pi\times 14.32\,$kHz \cite{Heavens1961}, although the level $6D_{3/2}$ has other decay channels, so that its total decay rate is in the range of $2.7\,$MHz~\cite{Safronova2016}. %The $6D_{3/2}$ level can be achieved by laser pumping from the state $6S_{1/2}$ at \textcolor{blue}{885} nm. 
In our model, we consider $|g\rangle\equiv|7P_{1/2}\rangle$ and $|e\rangle\equiv|6D_{3/2}\rangle$ as lower and upper (excited) states, respectively, and the atom is initially in the state $|e\rangle$, which can be prepared, e.g., by two-photon excitation from $|6S_{1/2}\rangle$ (see Fig.~\ref{fig2}a).

%The Casimir-Polder shift in Eq. \ref{delta_Ee} applies for both $\left| g \right\rangle$ and $|e\rangle$. It can be shown that for a single atom at a distance $h$ from surface, we have
%The frequencies of the Casimir-Polder shift given in \cite{Wylie1984} hold for both ground and excited states. 
The Casimir-Polder shift given in~\cite{Wylie1984} holds for both ground and excited states. It can be shown that for a single atom at a distance $h$ from a surface, we have %\Philippe{change $\delta_{CP}$ to $\delta$?}
\begin{equation}
    \delta=-\dfrac{C_3}{h^3}
    \label{delta_CP_ias}
\end{equation}
where $C_3=13.53\,\text{kHz}.\upmu\text{m}^3$ ($C_3=-100\,\text{kHz}.\upmu\text{m}^3$) for the $7P_{1/2}$ ($6D_{3/2}$) level \cite{Failache1999,Joao2023}. The decay rate $\Gamma_z$ has a similar dependence. 

In section \ref{sec_results}, we will consider the dipole orientation $\mathbf{\hat{d}}$ perpendicular to the surface for all atoms, i.e., $\mathbf{\hat{d}}=\mathbf{\hat{z}}$. For Cs and a sapphire surface, we checked that %\st{the direction of $\mathbf{\hat{d}}$} 
%\Philippe{
orienting all dipoles parallel to the surface (i.e., $\mathbf{\hat{d}}=\mathbf{\hat{x}}$) does not significantly alter the impact of the surface on the decay dynamics. We also checked that our simulations give $\delta=-(113.39\,\text{kHz}.\upmu\text{m}^3)/h^3$ for a single atom and sapphire, in agreement with Eq.\,\eqref{delta_CP_ias} where the global shift is $|C_{3tot}|=$%\st{13.53+100=}
$113.53\,\text{kHz}.\upmu\text{m}^3$. For interaction with metallic surfaces (ideal conductor; $\varepsilon=-\infty$), it is known that a single atom, whose dipole moment is oriented %\Philippe{whose dipole moment is oriented} 
perpendicularly to the surface, interacts stronger as compared to an atom placed parallel to the surface~\cite{Arruda2017,Arruda2020}. For very low distances $kh$ from the surface (typically $h\sim 10\,$nm), we have $\Gamma_z\approx 0$ ($\Gamma_z\approx 2\Gamma$) for $\mathbf{\hat{d}}\parallel\mathbf{\hat{z}}$ ($\mathbf{\hat{d}} \perp \mathbf{\hat{z}}$). We checked that our simulations reproduce this effect.

\begin{table}
    \caption{\label{tab:example} Values of the parameters used for the simulation of a single Cs atom interacting via its transition $6D_{3/2}\rightarrow 7P_{1/2}$ with %\Philippe{used for the simulation of a single Cs atom interacting via its transition $6D_{3/2}\rightarrow 7P_{1/2}$ with} 
    a sapphire surface. The quantities $\delta$ and $\Gamma_z$ were evaluated with Eqs.~\eqref{Vab_Gab} for $a=b$ and $G=G^\text{R}$ (see text).}
    \begin{ruledtabular}
    \begin{tabular}{llll}
        $h$ ($\mu$m) & $kh$ & $\delta/\Gamma$ & $(\Gamma+\Gamma_z)/\Gamma$\\
        0.100 & 0.05 & 5660 & 29235 \\
        0.500 & 0.25 & 63 & 176 \\
        1 & 0.5 & 9 & 14 \\
        2 & 1 & 1 & 1.074 \\
        3 & 1.5 & 0.24 & 0.61 \\
        4 & 2 & 0.009 & 0.73 \\
        5 & 2.5 & -0.06 & 0.9 \\
        10 & 5 & 0.005 & 0.98 \\
    \end{tabular}
    \label{tab1}
    \end{ruledtabular}
\end{table}

%\begin{table}
%    \caption{\label{tab:example} Values of the parameters of the simulation for a single Cs atom, transition $6$ D$_{3/2} \rightarrow 7$ P$_{1/2}$, close to a sapphire surface. The quantities $\delta$ and $\Gamma$ were evaluated with Eqs.~\eqref{Vab_Gab} for $a=b$ and $G=G^\text{R}$ (see text). \Philippe{(What is the meaning of negative decay rates?)}}
%    \begin{ruledtabular}
%    \begin{tabular}{llll}
%        $h$ ($\mu$m) & $kh$ & $\delta/\Gamma$ & $\Gamma_z/\Gamma$\\
%        0.100 & 0.05 & 5660 & 29234 \\
%        0.500 & 0.25 & 63 & 175 \\
%        1 & 0.5 & 9 & 13 \\
%        2 & 1 & 1 & 0.074 \\
%        3 & 1.5 & 0.24 & -0.39 \\
%        4 & 2 & 0.009 & -0.27 \\
%        5 & 2.5 & -0.06 & -0.1 \\
%        10 & 5 & 0.005 & -0.02 \\
%    \end{tabular}
%    \label{tab1}
%    \end{ruledtabular}
%\end{table}

Table\,\ref{tab1} shows some values of $h$ and the corresponding values of $kh$, shifts, and decay rates used in our simulations, for Cs and sapphire.

%\subsection{Cooperative effects for a line of atoms in free space}

%The aim of this paper is to see how cooperative effects are modified when a surface is placed close to an atomic system. In order to get a insight of the fluorescence emitted by a line of atoms, we analyse first $P(t)$ for $N$ atoms in free space, for several $kd$ values.

%Figure \ref{fig3} shows the fluorescence $P(t)$ for five atoms ($N=5$) in free space (no surface), for three different values of $kd$. When the atoms are far from each other ($kd=10$ and $kd=3$; dashed red and green curves), no cooperative effects are seen because the many-atom decay is close to the single atom decay. However, when the atoms are close to each other ($kd=1$), we see that $P(t)$ presents cooperative effects: superradiance (fast decay; $\Gamma^{(s)}t < 4$) and subradiance (slow decay; $\Gamma^{(s)}t > 4$). It is worth to mention that in~\cite{Jones2018} we have $kd=0.5$ for a line of $N=20$ atoms, where $\lambda=2.5$ $\mu$m for Sr atoms. Long-lived cooperative states were observed in free-space in this configuration~\cite{Olmos2013}. These results are similar to the ones studied by ????, where ???? \Philippe{Procurar artigos com simulacao e experimentos com linha de atomos}

%\section{Results and discussion}
%\textcolor{red}{
\section{ Results}
%}
\label{sec_results}

In what follows, we consider a line of $N=5$ Cs atoms interacting with a sapphire surface. %, unless otherwise mentioned. 
As already mentioned, the atoms are aligned perpendicularly to the surface, i.e., $\mathbf{\hat{d}}=\mathbf{\hat{z}}$. We checked that simulations with $N=20$, $50$ and $100$ atoms give the same results. %We also checked that simulations with all dipoles oriented parallel to surface ($\hat{d} // \hat{x}$) give very similar results. %\Philippe{I wonder whether the long range terms in the Green tensor are sufficient to spoil super-/subradiance. Do you observe the same behavior when you set the short range artificially to 0 in the simulations?}

%\textcolor{blue}{I ran simulations considering only the long-range terms, i.e., by setting
%\begin{eqnarray*}
  %  V_{ab}^0&=&\frac{3\Gamma}{4}\frac{\cos\kappa_{ab}}{\kappa_{ab}} \\
   % \Gamma_{ab}^0&=&\frac{3\Gamma}%{2}\frac{\sin\kappa_{ab}}{\kappa_{ab}}
%\end{eqnarray*}
%in Eqs. 10. This is the so-called scalar approximation. For the line of atoms interacting with surface, this indeed does not change at all, so the cooperative effects are suppressed. However, in free space the cooperative decay is different for $kd=1$ (atoms very close to each other), meaning that the short range terms are important when atoms are closer.
%}

\subsection{Suppression of the cooperative effects due to surface interactions}

\begin{figure}[h]
    \centering
    \includegraphics[scale=1]{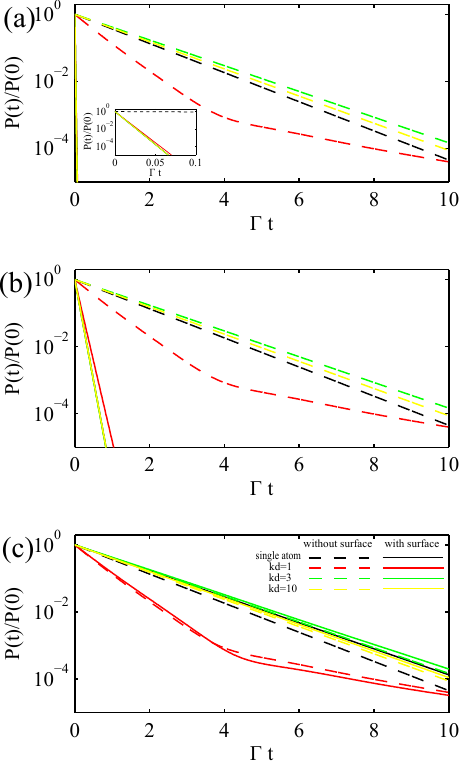}
    \caption{ Total fluorescence as a function of time for $\Delta=10\Gamma$ and (a) $kh=0.25$ ($h\approx 500\,$nm), (b) $kh=0.5$ ($h\approx 1\upmu$m),  and (c) $kh=2.5$ ($h\approx 5\,\upmu$m). Dashed (full) lines are for atoms without (with) sapphire surface, for: $N_\text{at}=1$ (black), $(N_\text{at},kd)=(5,10)$ (yellow), $(5,3)$ (green), and $(5,1)$ (red). Inset: zoom on panel (a).}
    \label{fig4}
\end{figure}

The main results of this paper are shown in Fig.~\ref{fig4}, for far detuned excitation. $P(t)$ is plotted for three values of $kd$, for atoms far from each other (solid green and yellow curves) and close to each other (solid red curves). The three panels (a), (b), and (c) correspond to atom-surface distances of $kh=0.25$, $kh=0.5$, and $kh=2.5$, respectively. %\Philippe{The three panels (a), (b), and (c) correspond to atom-surface distances of $kh=0.25$, $kh=0.5$, and $kh=2.5$, respectively.} 
The fluorescence of a single atom is plotted for comparison (black curves), as well as for atoms in free space, plotted as dashed lines. In Fig.~\ref{fig4}a, for atoms in free space (dashed lines), no cooperative effects are seen for $kd=3$ and $kd=10$, because the many-atom decay is close to the single atom decay (black dashed curve). However, when the atoms are close to each other ($kd=1$), we see that $P(t)$ presents cooperative effects: superradiance (fast decay; $\Gamma t<4$) and subradiance (slow decay; $\Gamma t>4$). It is worth mentioning that ref.~\cite{Jones2018} considers $kd=0.5$ for a line of $N=20$ atoms, where $\lambda=2.5\,\upmu$m for Sr atoms. Long-lived cooperative states were %\st{observed} 
studied theoretically in free-space in this configuration~\cite{Olmos2013}.

The impact of the surface consists in accelerating the decay, as seen in the inset of Fig.~\ref{fig4}a (solid lines) as almost-vertical curves, for atoms very close to the surface ($kh=0.25$ or $h\approx 500\,$nm). The data for a single atom and for $N=5$ for $kd=1,3,10$ almost collapse into a single curve. For a single atom, we have a high value of the surface decay rate, which is equal to $\Gamma_z=175\Gamma$ (see Table~\ref{tab1}). In~\cite{Failache1999}, for a hot vapor, it was observed that the atom loses its excitation to the surface modes. For $N>1$ atoms, we also have a fluorescence decay with a rate equal to the single-atom one. %This means that the cooperative effects are suppressed and the predominating decay is due to the Casimir-Polder interactions and induced by the coupling of the surface modes with the atoms.
This means that subradiance, the slow decay, is completely suppressed and %predominating decay is a very %\st{enhanced superradiance, the} enhanced decay due to the Casimir-Polder interactions and induced by the coupling of the surface modes with the atoms.
the decay rate is dominated by emission into the evanescent polariton modes of the surface.

Figures~\ref{fig4}b and \ref{fig4}c show the dependence of $P(t)$ on the distance $kh$ between the atoms and the surface. The free space data is displayed again for comparison. If the line of atoms is moved away from the surface (Fig.~\ref{fig4}b with $kh=0.5$), the surface interactions become weaker, and the decay slows down. For atoms very far from the surface (Fig.~\ref{fig4}c with $kh=2.5$), the surface effects disappear completely. In this case, we recover the original decays (solid and dashed curves coincide), and cooperative effects reappear. %\Philippe{cooperative effects reappear}.

% Fig 3: Fluo x t for several kd and (a) small and (b,c) large kh - sapphire - Delta=10

We checked that similar results are obtained for on-resonance illumination %atoms at resonance 
(i.e., $\Delta=0$), and when the %\st{polarization} 
direction of the incoming laser field is aligned with the atoms (i.e., %\st{$\mathbf{\hat{d}}=\mathbf{\hat{y}}$} 
incident laser along the $y$ axis). In free space, subradiance at resonance has its weight increased, as %\st{observed} 
studied theoretically in \cite{Guerin2017, Araujo2018}, due to a larger occupation of the subradiant modes in steady state before switching off the incident driving~\cite{Guerin2017}. The same occurs for laser excitation parallel to the atom line, because in this geometry the optical depth $b_0$ of the system is larger, and cooperative decay rates depend on $b_0$ \cite{Guerin2016,Araujo2016,Roof2016}. However, similar to the far-detuned case displayed in Fig.~\ref{fig4}, cooperative effects for $\Delta=0$ and parallel excitation are completely absent and they are replaced by a very fast decay caused by atom-surface interactions at small $kh$.

\subsection{Impact of the surface resonance in the atom coupling}

The Casimir-Polder effects discussed in the previous subsections are dominated by the fact that the surface is resonant with the atomic transition, i.e., $\lambda_\text{s}\approx\lambda$ ($\varepsilon=-1$). Surfaces such as glass and metal present resonances far from the atomic transitions currently used in experiments. As an example, an Ag surface presents a plasmonic resonance around $3.64$ eV~\cite{Jones2018}, whereas earth and earth alkali atoms have their main transitions below this value, e.g., $0.5\,$eV for the Sr transition $^3P_0\rightarrow\,^3D_1$, $1.6\,$eV for the Rb $D_2$ line, $2.1\,$eV for the Na $D_2$ line, or $2.7\,$eV for the Sr transition $^1S_0\rightarrow\,^1P_1$.

% Fig 7: Fluo x t for several lambda
\begin{figure}[h]
    \centering
    \includegraphics[scale=1]{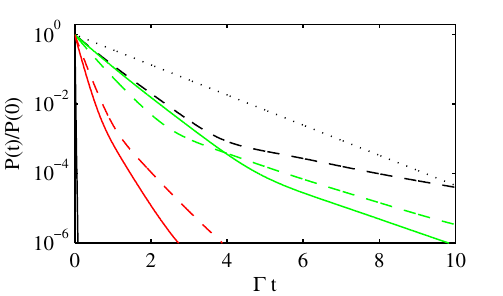}
    \caption{Emitted fluorescence $P(t)$ for $N=5$ atoms, $kd=1$ and $\Delta=10\Gamma$, interacting with surface  for the following atomic wavelengths: $\lambda=8.15\,\upmu$m (dashed light green), $10.15\,\upmu$m (full light green), $12.15\,\upmu$m (full black), $14.15\,\upmu$m (full red) and $16.15\,\upmu$m (dashed red). These $\lambda$ give, respectively, the sapphire dielectric constants $\varepsilon=1.8+0.015i$, $\varepsilon=0.78+0.040i$, $\varepsilon=-0.95+0.11i$, $\varepsilon=-4.6+0.43i$ and $\varepsilon=-12+4.0i$. Dotted and dashed black lines: decay in free space for $N=1$ (dotted) and $N=5$ (dashed).}
    \label{fig7}
\end{figure}

In order to illustrate the impact of the surface resonance, Fig.~\ref{fig7} displays the fluorescence $P(t)$ emitted by the atoms interacting with sapphire for four different atomic wavelengths $\lambda$, below and above the resonant wavelength, $\lambda_\text{s}=12.15\,\upmu$m (we have evaluated the value of $\varepsilon$ for sapphire for each $\lambda$). Data without surface are displayed for comparison. When the surface is not resonant with the transition wavelength, we have $\varepsilon\neq -1$, and this retrieves some cooperative modes, meaning that the atom-surface coupling is not strong any more. Some remaining surface effects are due to evanescent surface modes, but not polariton ones. %However
On the other hand, for $\varepsilon=-1$, the cooperative decay is extinguished. We checked that similar results are obtained when fixing $\lambda_\text{s}=12.15\,\upmu$m but using values of $\varepsilon$ for glass (which presents no resonance around $12\,\upmu$m), an ideal metallic surface (where $\varepsilon=-\infty$) and an Ag surface. 

\begin{figure}[h]
    \centering
    \includegraphics[scale=1]{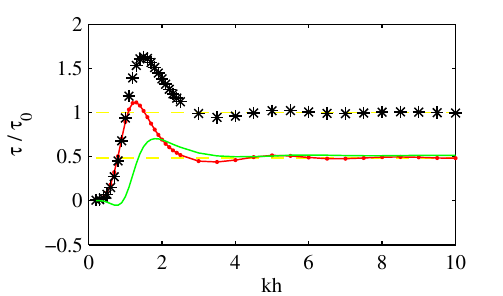}
    \caption{ Fitted time decay rates $\tau/\tau_0$ for several distances $kh$ from a sapphire surface at $\Delta=10\Gamma$, for a single atom (black stars) and for $N=5$ atoms separated by $kd=1$ (red dots). The dashed yellow lines at $\tau/\tau_0=1$ and $\tau/\tau_0=0.49$ represent, respectively, the free-surface single atom decay and the free-surface five-atom decay. The green full curve is the subtraction of the red-dots curve from the black-stars curve.}
    \label{fig5}
\end{figure}

In order to study the transition between the Casimir-Polder fast decay and the standard cooperative superradiant decay, we calculated the decay rate $\tau$ as a function the atom-distance $kh$, for a fixed number of atoms and atom separation $kd$. The results are displayed in Fig.\,\ref{fig5}. We also display the data for a single atom for comparison. The initial time decay rates $\tau$ were extracted from an exponential fit $P(t)/P(0)\propto e^{-t/\tau}$ of the respective fluorescences in the interval $t/\tau_0\in[0,1]$, where $\tau_0=1/\Gamma$ is the single-atom time decay rate in free space. For the line of atoms (red dots), $\tau$ increases until a maximum, reached for $kh=1.3$. Then, it decreases and oscillates until surface effects disappear and superradiance in free space is recovered (yellow dashed line at $\tau/\tau_0=0.49$). This far field oscillatory behavior of the single-atom emission rate is a well-known effect~\cite{Wylie1984}, also predicted in our simulations.

% Fig 5: tau_fit for several kh

On the other hand, the behavior of the single-atom decay (black stars in Fig.\,\ref{fig5}) is purely due to the change of the polariton decay rate $\Gamma_z$ with $kh$, since cooperative effects are absent. The single-atom decay curve has a similar shape as the multi-atom decay curve (red dots), and it reaches its maximum at $kh=1.5$. The full light green curve represents the subtraction of the multi-atoms decay curve from the single-atom one. Close to the surface (small $kh$) the single- atom and the multi-atom decay rates are the same, pointing to the absence of collective effects. Far from the surface (large $kh$) the difference curve shows a superradiantly enhanced decay from $\tau/\tau_0=0.7$ down to $0.5$.

\section{Discussion and outlook}
\label{sec_discussion}

The coupled-dipoles model had important successes in the last two decades~\cite{Bienaime2010,Guerin2016, Araujo2016} and turned into a standard tool for describing cooperative effects, such as super- and superradiance~\cite{Araujo2016, Roof2016}, cooperative shifts~\cite{Roof2016} and even atomic correlations in certain regimes~\cite{Piovella2022}. %\textcolor{gray}{In some situations surfaces may not be beneficial at all. Maybe for metamaterials, where the local distance of surface is reduced or shaped in such a way that atom-light interaction has not a dominant absorptive character. In any case, it is important to have at hand a working theoretical tool, and that is what we provide.}
Based on this model, throughout this work we looked at the total emitted fluorescence, which is sensible because of the non-directional character of subradiance~\cite{Bienaime2012}. %On the other hand, superradiant decay \textcolor{blue}{is anisotropic}~\cite{Araujo2016}.% \Philippe{is anisotropic} \st{rate depends on direction}~\cite{Araujo2016}.%\st{, although it can be observed in all directions. Realistic geometries like a sphere would introduce different frequency surface shifts for each atom, but we expect no qualitative changes.}

Experiments can be performed with cold optical lattices or even thermal clouds of atoms. The presence of super- and subradiance has been demonstrated theoretically and experimentally without nearby surfaces using Rb atomic clouds with temperatures between $\sim50\,\upmu$K and $\sim10\,$mK ~\cite{Weiss2019}. On the other hand, experiments with atoms close to surfaces are challenging, because the precise control of their distance to the surface requires extremely low temperatures, e.g., below the critical temperature for Bose-Einstein condensation~\cite{Bender2010}.

Resonances between sapphire and Cs via the transition $6D_{3/2}\rightarrow 7P_{1/2}$ at $12.15\,\upmu$m were extensively studied for hot vapors in the $1990$s and $2000$s~\cite{Fichet1995, Failache1999, Failache2002, Laliotis2014, Laliotis2015, Joao2023}. We used this transition in our simulations as a showcase, leaving out, for simplicity, the existence of faster decay channels such as the decay $6D_{3/2}\rightarrow 6P_{1/2}$ at $876$\,nm which, in reality, can have an important impact on the dynamics. In this respect, the first resonances of the alkalis (D1 or D2 lines) or Sr would be much better candidates for exploring collective fluorescence effects close to resonant surfaces, because they represent real two level systems without additional decay channels. Unfortunately, most dielectrics do not exhibit surface resonances at optical or near infrared frequencies.

A way of overcoming this limitation can be the use of nanofabricated metasurfaces engineered such as to tune resonances to predefined frequencies~\cite{Aljunid2016, Chan2018, Chan2019}. An experimental demonstration of the coupling between the D2-line in Cs (see Fig.~\ref{fig2}a) and a metasurface whose resonance was tuned close to $852$\,nm has already been reported~\cite{Chan2018}. Assuming an effective dielectric constant for the metamaterial and using $852$\,nm as the transition wavelength, we find qualitatively the same decay behavior as in Fig.\,\ref{fig4}a.

In order to observe polariton-induced superradiant decay, one possibility would be to drive Cs atoms on their D2 line with a laser pulse, as done in~\cite{Guerin2016,Araujo2016, Roof2016}, directly comparing two configurations: close to a glass surface and close to a metamaterial surface. For glass, which has no polariton mode near the D2 line, %\st{$6D_{3/2}\rightarrow 7P_{1/2}$ transition},
the fluorescence is expected to exhibit standard super- and subradiant decays~\cite{Araujo2016,Roof2016}. For a suitable metamaterial with a resonance near the D2 line, induced by the surface polariton, the atoms should quickly decay to the ground state. %Most of the released fluorescence will be absorbed by the resonant surface; however, for a thin metamaterial surface, some fluorescence presenting signatures of superfast polariton-induced decay will not be absorbed, which then can be measured.

%\textcolor{red}{\section{Conclusion and outlook}}
\section{Conclusion}
\label{sec_conclusion}

In summary, we have studied the impact of resonant surfaces on the dynamics of nearby atoms in the low-excitation regime. At the example of a line of Cs atoms close to a sapphire surface exhibiting a resonance close to an atomic transition, we found suppression of cooperative effects, contrasting with a reported enhancement of the Casimir-Polder effect for strong excitations~\cite{Fuchs2018,Sinha2018}. In the absence of surface polaritons close to an atomic transition, cooperative effects are observed, although they are slightly modified due to the interaction with evanescent vacuum modes that are still present far from resonance. We provide a working theoretical tool taking into account the successes of both, the coupled-dipoles model and atom-surface interactions between Cs and sapphire and suggested an experiment relying on cold Cs atoms interacting with a metamaterial surface.

Useful applications, e.g., in quantum information~\cite{Santos2022} and metrology~\cite{Pineiro2020} rely on the possibility to control cooperative effects. An interesting example is the superradiant laser, where the collective coupling of an atomic sample to a dissipative mode (e.g., an optical cavity operated in the 'bad cavity' limit) leads to global synchronization of the atomic dipoles~\cite{Bohnet2012}. The interaction with surfaces may provide a lever to handle this control. %\textcolor{red}{Concerning cavities with many atoms, cavity has to have a small size, in the order of the laser wavelength, and a theoretical model should include several dipole images for the atoms.}

%It is important to understand the impact of surfaces on collective dynamics in various circumstances. This work shows that the vicinity of polariton resonances can have a devastating effect on cooperativity. Although performed in the weak excitation limit, we believe that our results will contribute to the understanding of cooperative atom-surface coupling in the presence of many excitations.

%\st{Controlling of cooperative effects, specially subradiance, has prospecting applications in quantum information}~\cite{Santos2022} and metrology~\cite{Orioli2020}. \st{We believe this work demonstrates that the coupling between cooperative and surface modes, at least for surface resonances and in the weak excitation regime, can contribute to the understanding of this coupling when many excitations are present.}

\section*{Acknowledgements}

M.O.A. and J.C.d.A.C. thank UFPE for some financial
support. Ph.W.C. received support from FAPESP (Grant
No. 2022/00261-8), from CAPES-COFECUB (Grant No.
88887.130197/2017-01), and from CNPq-SFSN (Grant No.
402660/2019-6).

%M.O.A. and J.C.de A.C. contributed equally to this work.

%\textcolor{blue}{M.O.A. contributed to planning of the project, simulations and analytical calculations. J.C. de A.C. contributed to the conception and data interpretation. Ph.W.C. contributed to analytical calculations, discussion and interpretations. A.L. contributed with data interpretations.}

%%%%%%%%%%%%%%%%%%%%%%%%%%%%%%%%%%%%%%%%%%%%

% \bibliography{refs_papers}

% ----- OR ---------------------------------

\bibliographystyle{apsrev4-1}

\end{document}